# Snap-Through Thermomechanical Metamaterials for High-Performance Thermal Rectification

Qinyun Ding[1], Yuhao Wang[1], Guanqing Xiong[1], Wei Chen[1], Ying Chen[2], Zhaoguang Wang[1], Arup Neogi[3], Jaehyung Ju[1]*

1. UM-SJTU Joint Institute, Shanghai Jiao Tong University, Shanghai 200240, Shanghai, China.

2. School of Aerospace Engineering, Xiamen University, Xiamen 361000, Fujian, China.

3. Institute of Fundamental and Frontier Sciences, University of Electronic Science and Technology of China, Chengdu 610054, Sichuan, China.

*Corresponding author: Jaehyung Ju (jaehyung.ju@sjtu.edu.cn)

**Abstract:**

Thermal diodes that enable directional heat transport are essential for advanced thermal management in microelectronics, energy systems, and thermal logic devices. However, existing designs based on phase-change materials, nanostructures, or interfacial engineering suffer from limited rectification performance, configurational inflexibility, and poor scalability. Here, we present a thermomechanical metamaterial-based thermal diode that combines temperature-responsive actuation with structural bistability to achieve high-efficiency, nonreciprocal thermal transport. The device integrates shape memory alloy (SMA) springs with pre-buckled copper strips that undergo snap-through transitions in response to thermal gradients. This reconfiguration enables contact-based conduction in the forward mode and suppresses reverse heat flow via radiative isolation. We develop a coupled analytical model combining Euler–Bernoulli beam theory and a thermal resistance network, and validate the system through finite element (FE) simulations and experiments. The device achieves a thermal rectification ratio exceeding 900, with robust cycling stability and structural integrity. A modular stacking strategy further enhances scalability without compromising performance. This work establishes a new design framework for high-performance, passive thermal rectifiers that bridge mechanical metamaterials and advanced thermal engineering.

# 1. Introduction

Efficient thermal management is essential for a broad range of technologies, including microelectronics[1], battery systems[2,3], thermal logic devices[4-8], and spacecraft thermal regulation[9]. In these systems, directional heat transport—where heat preferentially flows in one direction while being suppressed in the opposite—enables enhanced energy efficiency, protection against thermal backflow, and novel functionalities in thermal control. Devices that realize this behavior are called thermal diodes or thermal rectifiers[10,11].

Conventional thermal diodes exploit temperature-dependent material properties such as phase transitions[12,13], interfacial asymmetry in graphene heterostructures[14], or phonon scattering in silicon-based nanostructures[15,16]. While these designs provide proof-of-concept functionality, they are limited by several key challenges. Phase-change and silicon-based devices often achieve low rectification ratios (<11)[12,15,16]; droplet-based systems suffer from performance degradation due to flooding[13]; and graphene-based approaches lack geometric tunability and are often constrained to simulated configurations[14].

To overcome these limitations, thermomechanical metamaterials[17-19] offer a new design paradigm by coupling mechanical deformation with thermal responsiveness. Recent demonstrations include shape memory alloy (SMA)-based thermal switches activated by electric current[20], kirigami-inspired thermal regulators[21,22], and mechanically actuated thermal transistors for logic operations[23]. Other developments in thermal metamaterials, such as transformation-based thermal illusions[24] and transient rectification in macroscale systems[25], have further expanded the functional landscape of directional heat control. While these studies have advanced specific functionalities or conceptual designs, an integrated system that offers predictive modeling, experimental validation, and high steady-state rectification performance remains lacking—a gap this work seeks to fill.

Recent demonstrations include SMA-based thermal switches activated by electric current[20], kirigami-inspired thermal regulators[21,22], and mechanically actuated thermal transistors for logic operations[23]. While these studies have advanced specific functionalities, an integrated system that offers predictive modeling, experimental validation, and high rectification performance remains lacking.

In this study, we introduce a metamaterial-based thermal diode that combines geometrically nonlinear bistability with temperature-responsive actuation to achieve directional heat transport with unprecedented efficiency. The device consists of engineered copper strips that exhibit snap-through transitions between two stable configurations, triggered by localized heating of embedded SMA springs. This thermally induced actuation reconfigures the structure to enable low-resistance conduction in the forward mode via direct contact, while reverse heat transfer is suppressed through engineered air gaps that shift the transport mechanism to radiation.

By integrating snapping of a column-buckled beam theory with a thermal resistance network model, and validating the system through finite element (FE) simulations and experiments, we demonstrate a thermal rectification ratio exceeding 900. The design offers robust performance under thermal cycling and is inherently scalable via a modular stacking architecture. This work establishes a new class of passive, reconfigurable thermal diodes, bridging thermomechanical metamaterials with advanced thermal management strategies, and opening pathways for multifunctional heat regulation in next-generation energy and computing systems.

## 2. Design of thermal diodes

SMAs exhibit the shape memory effect (SME)—a thermally driven phase transformation that enables recovery of a pre-deformed shape. At low temperatures, SMAs exist in the martensitic phase, which can be mechanically detwinned to produce macroscopic deformation. Upon heating above the austenite finish temperature ($T_A^F$), the material transforms into the high-symmetry austenite phase, restoring its original shape. This reversible transformation is diffusionless and driven by atomic lattice-level shear, making SMAs ideal for thermal actuators[26-29].

We harness this thermally driven actuation to design a thermomechanical metamaterial that functions as a thermal diode (Figure 1a). The device consists of SMA springs and copper strip beams, as illustrated in Figure 1b. When exposed to heat, the SMA springs, connected via rotational hinges to the sidewalls of the base, transform from detwinned martensite to austenite, generating force that pushes the hinges at the ends of the copper beam. This actuation causes the snapping of a column-buckled copper strip, as shown in Figure 1c, creating or removing thermal contact.

In the backward mode ($T_1 < T_2$, Figure 1d), the SMA springs remain in the martensitic phase and do not actuate, maintaining an open gap between the central block and the conductive block. This configuration impedes thermal conduction across the gap. In the forward mode ($T_1 > T_2$, Figure 1c), the SMA springs are heated above $T_A^F$, triggering the phase transformation and extension. The resulting couple moments at the hinges cause the buckled copper strip to snap into contact, allowing the central block to press against the conductive block. This establishes a heat-conducting path via the copper strip, enabling directional heat flow from the base to the block.

By coupling SMA actuation with snap-through instability of the copper strips, our design achieves nonreciprocal thermal transport. This property allows the system to selectively permit or block heat flow depending on the thermal gradient, which is valuable for thermal management and regulation applications. Our device experimentally demonstrates a record-high thermal rectification ratio of ~936, outperforming previously reported thermal diodes.

The key design principle lies in integrating the temperature-induced deformation of SMAs with the mechanical bistability of copper strips to achieve direction-dependent thermal response. To control the copper strip deformation, we apply an energy minimization approach to derive unknown parameters in the expression[30,31] (Equations (1)–(5)) The thermal performance is quantified via the thermal rectification ratio, accounting for both forward conduction and backward radiative resistance, as described in Section 5.

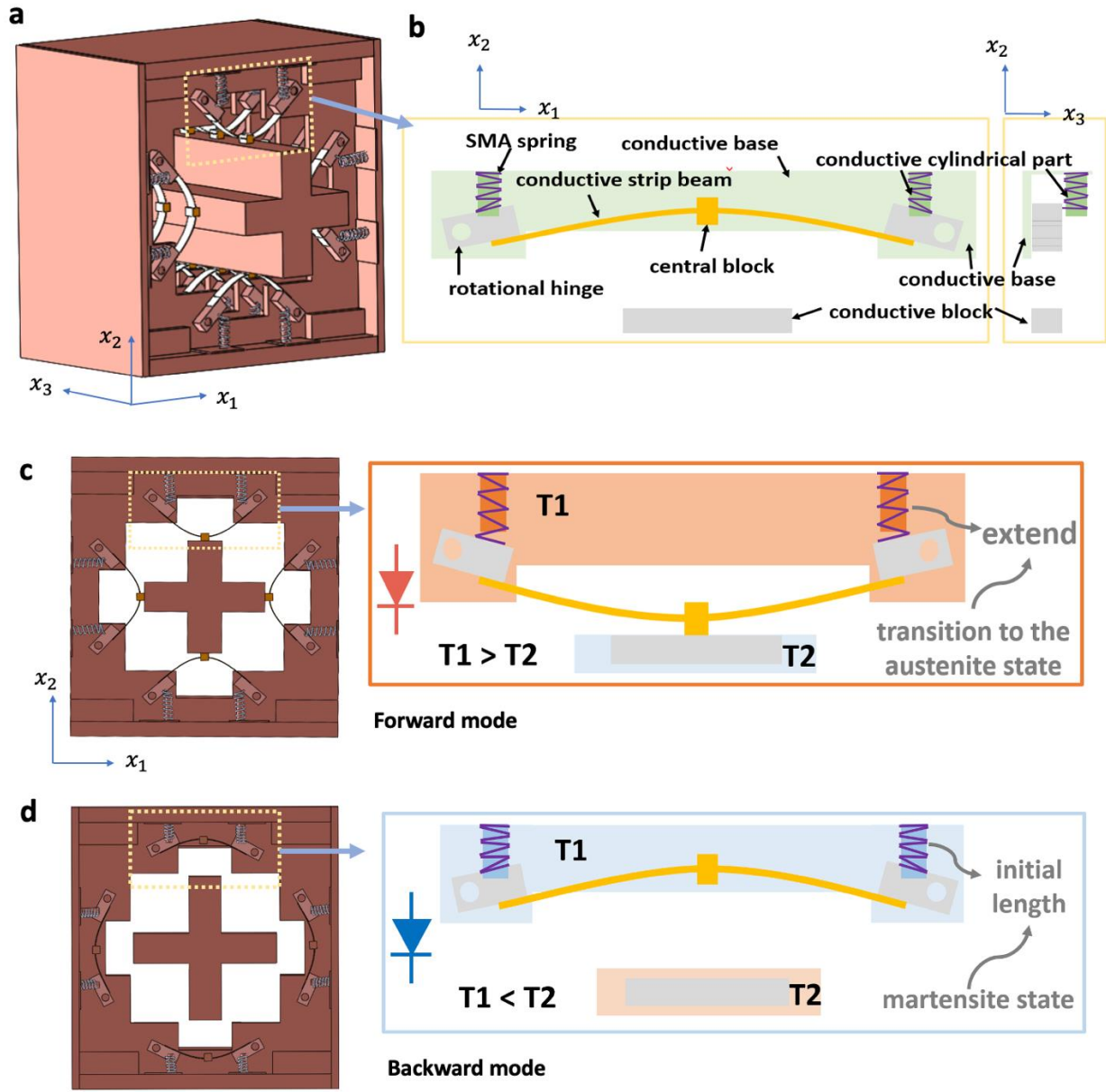

*Figure 1. (a) Oblique pictorial view of the thermomechanical metamaterial thermal diode. (b) Schematic of a single thermal diode showing SMA spring actuation and copper strip bistability. (c) The device operating in forward mode: SMA springs transform into austenite, actuate the hinges, and snap the copper strip into contact, enabling directional heat transfer. (d) The device operating in forward mode. (d) The device operating in backward mode: SMA springs remain in the martensitic phase, and a gap inhibits heat conduction.*

We develop an analytical model to describe the snap-through behavior of pre-compressed copper strips, actuated by thermally triggered SMA springs. Unlike Cleary et al.[32] and Zou et al.[30], who used fixed–fixed boundary conditions to model post-buckling shapes, we apply pinned–pinned boundary conditions, which more accurately reflect the moment-driven actuation by rotational hinges coupled with SMA springs (Figures 2a and 2b).

Thermal actuation arises as the SMA springs transform into the austenitic phase and extend, generating bending moments at the beam ends. As shown in Figure 2b, the moment is calculated as the product of the SMA spring force and the moment arm — the distance between the hinge center and the spring's line of action. Heating occurs via conductive cylinders embedded in the SMA springs, which are connected to a thermally conductive base (Figures 1b, 2b). The extension generates moments at both ends of the copper strip, initiating a snap-through transition to a new equilibrium configuration (Figure 2b).

We employ the Euler–Bernoulli beam theory to model the pre-compressed, pinned–pinned beam. The governing differential equation is:

$$EI\frac{d^4w}{dx^4} + P\frac{d^2w}{dx^2} = 0 \tag{1}$$

where $EI$ is the bending stiffness, $P$ is the axial pre-compression force, and $w(x)$ is the transverse deflection. Eq. (1) can be illustrated as:

$$\frac{d^4w}{dx^4} + \alpha^2\frac{d^2w}{dx^2} = 0 \tag{2}$$

where $\alpha = \sqrt{P/EI}$. Solving Eq. (2) under symmetric boundary conditions $w(0) = w(L) = 0$ and $M(0) = M(L) = 0$, where $L$ denotes the horizontal length after pre-compression and $M$ is the bending moment (Figures 2a and 2b). We express the deflection $w(x)$ using a superposition of the first and third buckling modes:

$$w(x) = A_1 \sin\left(\frac{\pi x}{L}\right) + A_3 \sin\left(\frac{3\pi x}{L}\right) \tag{3}$$

where $A_1$ and $A_3$ are the modal amplitudes. The selection of these two modes captures the symmetric deformation observed during snap-through, as illustrated in Figure 2b.

We calculate the total potential energy $\Pi$ of the system, consisting of Compression energy $U_C$, Bending energy $U_B$, and External work by moments $W_M$.

The total potential energy $\Pi$ is formulated by:

$$\Pi = U_C + U_B - W_M \tag{4}$$

The modal amplitudes $A_1$ and $A_3$ are determined by minimizing $\Pi$ with respect to both variables. We solve the system:

$$\frac{\partial \Pi}{\partial A_1} = 0, \quad \frac{\partial \Pi}{\partial A_3} = 0 \tag{5}$$

However, an additional unknown — the applied moment $M$ — results in three unknowns with only two equations. To resolve this, we treat $M$ as an iterative parameter and solve the system using a numerical algorithm (Figure 2c). The MATLAB `vpasolve` function is used to find real-valued solutions, terminating the iteration when the moment returns to zero or a solution is not found.

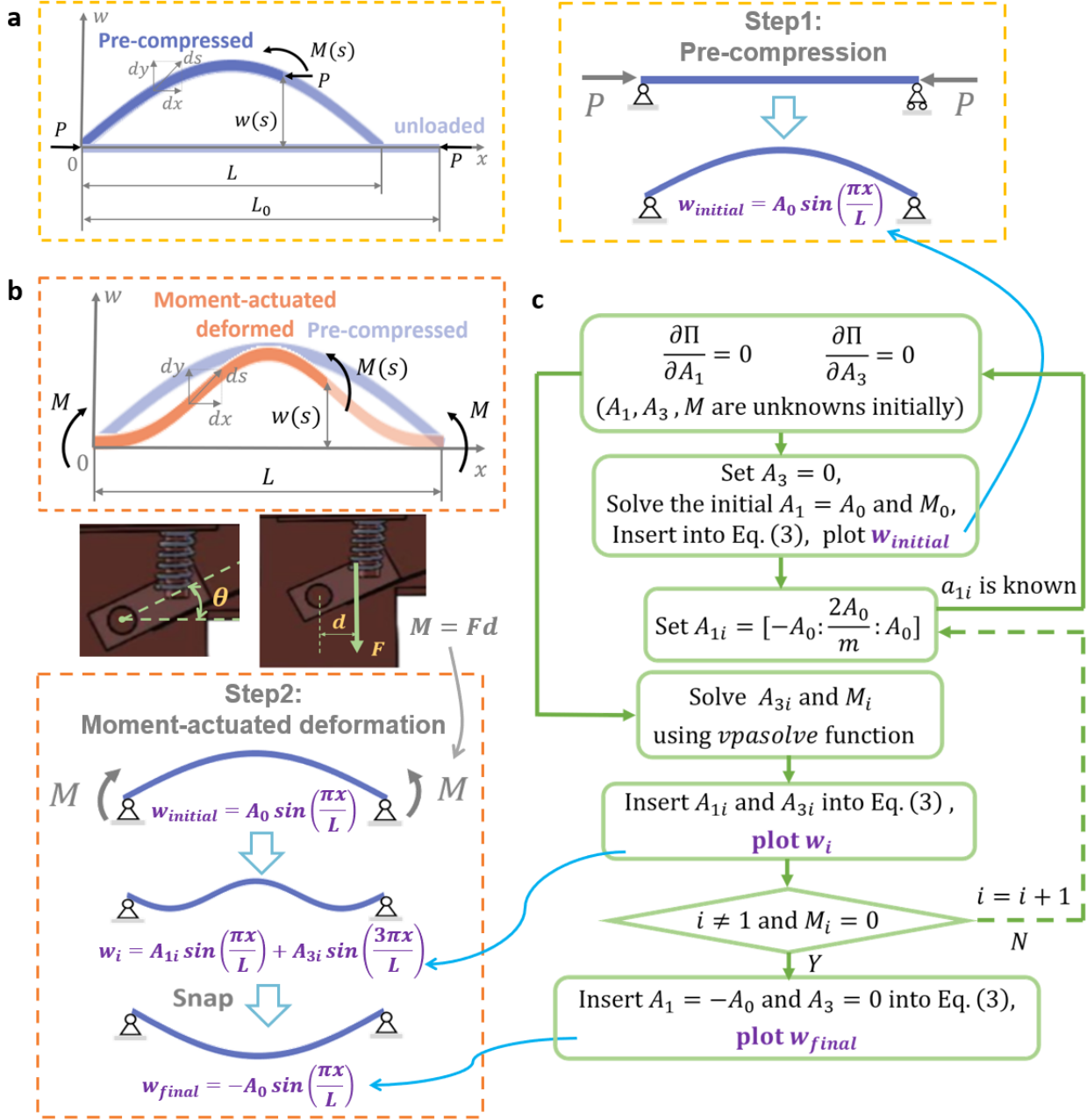

Figure 2. Mechanics of SMA-actuated snapping in copper strips. (a) Geometry and actuation setup for pinned–pinned pre-compressed copper strip showing initial buckling configuration. First step of deformation process: Initial buckling from axial compression. (b) Deformed shape represented by the superposition of the first and third buckling modes. Definition of rotational angle $\vartheta$ and bending moment $M$ induced by SMA spring extension. Second step of deformation process: Snap-through triggered by bending moments from thermally actuated SMA springs. (c) Flowchart of the numerical iteration algorithm used to solve modal amplitudes under moment constraints by minimizing the total potential energy. Here, $P$ denotes the axial pre-compression force, $M$ is the input moment applied by the SMA actuator, $M(s)$ is the internal bending moment, $w(s)$ is the transverse deflection of the beam (orthogonal to the x-axis), $L$ is the pre-compressed length along x-axis, $A_1$ and $A_3$ are the modal amplitudes, and $\Pi$ is the total potential energy.

## 3. Model Validation and Thermomechanical Model Establishment

We validate the analytical model through FE simulations and experimental measurements. As shown in Figure 3a, the predicted moment–deflection response exhibits excellent agreement with FE simulations, confirming the model's ability to capture the mechanical behavior of the bistable system. Figure 3b presents the central deflection of the buckled beam as a function of temperature, demonstrating a strong correlation between theoretical predictions and experimental observations. This agreement is derived from a coupled thermomechanical model that integrates the deflection–rotation of a copper strip obtained via iterative numerical analysis (Figure 2c) and a temperature–rotation constitutive relation (see Section S2, Supplementary Information). Notably, the sharp decrease in rotational angle near 44 °C corresponds closely to the onset of the SMA's austenitic transformation at 46 °C, while the transition completes near 50 °C, consistent with the austenite finish temperature. These results confirm that the observed snap-through transition is driven by SMA actuation and associated moment generation.

Figure 3c compares the beam's deformed configurations across multiple stages of actuation. The analytical and FE results show consistent shapes, while the experimental data reveal minor asymmetries attributed to fabrication error and nonuniform heating. For effective thermal diode operation, the post-snap deflection must bring the copper strip into contact with the heat sink block. This condition necessitates that the snap-through occurs before the SMA reaches full austenitic transformation (i.e., below $A_f$). If the actuation is delayed beyond this point, the beam may stabilize in an intermediate configuration (as illustrated in the middle state of Figure 3c), preventing closure of the thermal gap and compromising thermal conduction. Additionally, while the analytical and FE models predict symmetric deformation, the experimental observations in Figure 3c exhibit small but systematic deviations: the initial state is often skewed toward the right, whereas the final snapped configuration tends to shift left. This asymmetry arises from two distinct sources: initial misalignments introduced during manual assembly account for the rightward bias in the pre-compressed state, while the final state's leftward deviation results from uneven thermal actuation caused by asymmetric heating and slight differences in SMA spring response.

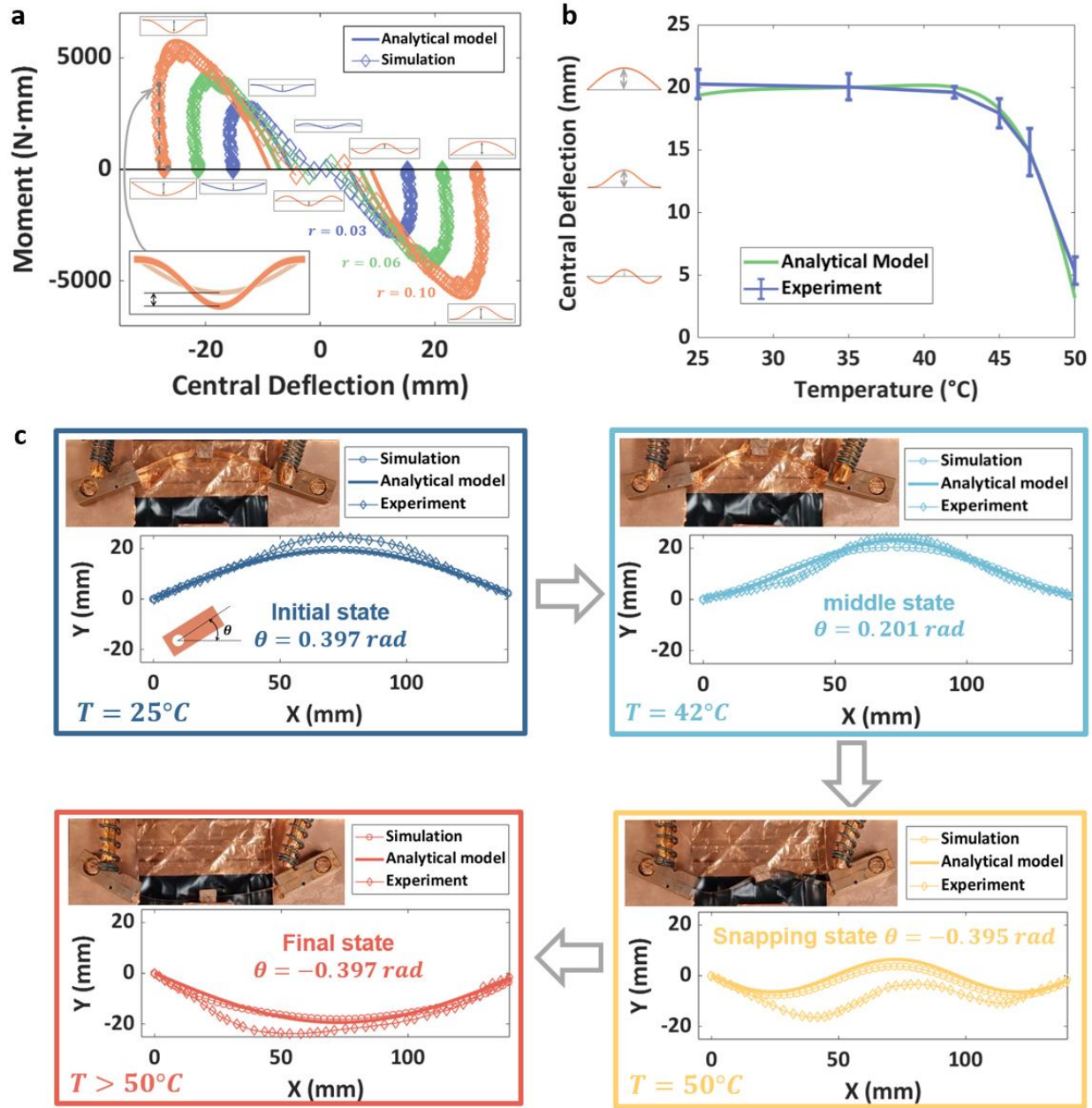

*Figure 3. Validation of the analytical model through FE simulations and experiments. (a) Moment–central deflection response predicted by the analytical model and FE simulations. Two symmetric deflection branches appear due to mirror-symmetric buckling paths. (b) Central deflection of the buckled beam as a function of temperature. The sharp drop near 44 °C corresponds to the SMA's austenite start temperature. (c) Evolution of the beam shape during snapping: Initial state (pre-compressed), Middle state (partial actuation), Snapping state (transition), and Final state (post-snap), where θ represents the rotational angle of the left hinge. Experimental deviations from symmetry in snapping and the final state arise from fabrication errors and heating inconsistencies.*

## 4. Phase Map for Design Optimization and Critical Temperature Estimation

We use the validated model to develop a design phase map that links geometric parameters (e.g., compression ratio $r$ and spring offset $d$) to the critical snapping temperature (Figure 4a). Increasing either parameter raises the required spring extension and thus the critical temperature. The snapping must occur within the SMA's transformation range (e.g., 46–64 °C). Figure 4b provides the maximum achievable beam deflection $w_{max}$ for given strip thicknesses and geometries, ensuring thermal contact. Figure 4c presents a flowchart for selecting SMA springs, configuring geometry, and ensuring sufficient beam deflection to meet the thermal diode requirement. These parameters are then linked to device performance via the rectification ratio shown in Figure 6a.

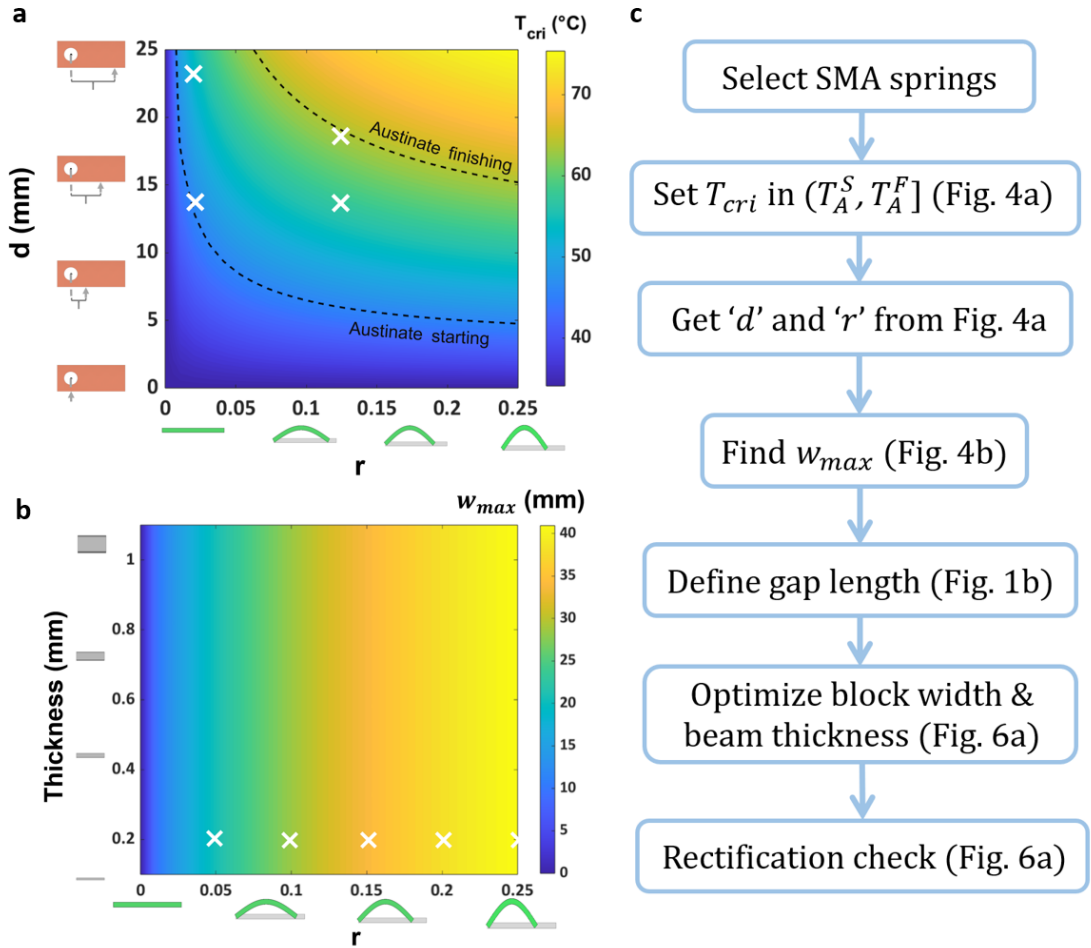

Figure 4. Design maps and methodology for thermal diode optimization. (a) Critical snapping temperature $T_{cri}$ as a function of two geometric parameters (offset $d$ and compression ratio $r = (L_0 - L)/L_0$), where $L_0$ is the initial length of the straight beam, and $L$ is the pre-compressed length along x-axis), showing constraints set by SMA transformation temperatures. (b) Maximum central deflection $w_{max}$ of the copper strip for various strip thicknesses and geometries, ensuring sufficient travel for thermal contact. (c) Design flowchart linking SMA selection, geometric tuning, and performance validation based on actuation temperature range and thermal rectification requirements, where $T_A^S$ is the austenite start temperature and $T_A^F$ is the austenite finish temperature.

## 5. Thermal Transport Modeling

Figures 5a and 5b show the temperature distributions in the forward and backward modes, as obtained from FE simulations. Because the thermal gradient is confined primarily to the conduction path, the thermal diode can be accurately modeled using a one-dimensional thermal resistance network. Equivalent thermal circuit diagrams for both operating modes are provided in Figure 5c.

In the forward mode, the exterior region (Region I) is at a higher temperature than the interior region (Region II), for example, $T_I = 333\text{K}$ and $T_{II} = 318\text{K}$. Heat flows from the conductive base through the rotational copper hinges and the buckled copper strip to the central copper block. The final step in this path is a contact interface between the central block and the underlying conductive base.

The total thermal resistance in the forward mode is expressed as:

$$R_f = \frac{1}{2}\sum_{i=1}^{2}(R_i + R_{c,i}) + R_3 + R_{c,3} = \frac{1}{2}\sum_{i=1}^{2}\left(\frac{L_i}{kA_i} + R_{c,i}\right) + \frac{L_3}{kA_3} + R_{c,3}, \qquad (6)$$

where:

- $R_i$: conduction resistances (hinges, strips, copper base, etc.) (Figure 5d),
- $R_{c,i} = \frac{R''}{A_{c,i}}$: thermal contact resistance (Figure 5d),
- $R'' = 1.7 \times 10^{-3} \text{ m}^2 \cdot \text{K/W}$: thermal contact resistance parameter,
- $A_{c,i}$: contact area between the central and conductive blocks (Figure 5d).
- $k = 400 \text{ W/m} \cdot \text{K}$: thermal conductivity of copper,
- $L_i$ and $A_i$: length and cross-sectional area of the $i^{th}$ copper segment (rotational hinges, copper strip, base),

This formulation captures both the conductive and interface-based contributions to heat transfer.

In the backward mode, the temperature gradient reverses: $T_{II} > T_I$, and the copper strip remains disconnected from the conductive block because SMA does not actuate at this condition. As a result, conduction is suppressed, and the dominant mode of heat transfer is only a thermal radiation across a micro-gap.

The total thermal resistance in this mode is:

$$R_b = \frac{1}{\epsilon \sigma A_r (T_{II}^2 + T_I^2)(T_{II} + T_I)}, \qquad (7)$$

where:

- $\epsilon = 0.055$: effective surface emissivity,
- $\sigma = 5.67 \times 10^{-8} \text{ W/m}^2\text{K}^4$: Stefan–Boltzmann constant,
- $A_r$: radiating surface area (see Supplemental Information for values).

Equation (7) is derived under two assumptions: (i) a unity view factor, i.e., all radiation emitted by one surface is intercepted by the opposing surface, and (ii) gray-body behavior between the copper surfaces – a constant emissivity independent of wavelength.

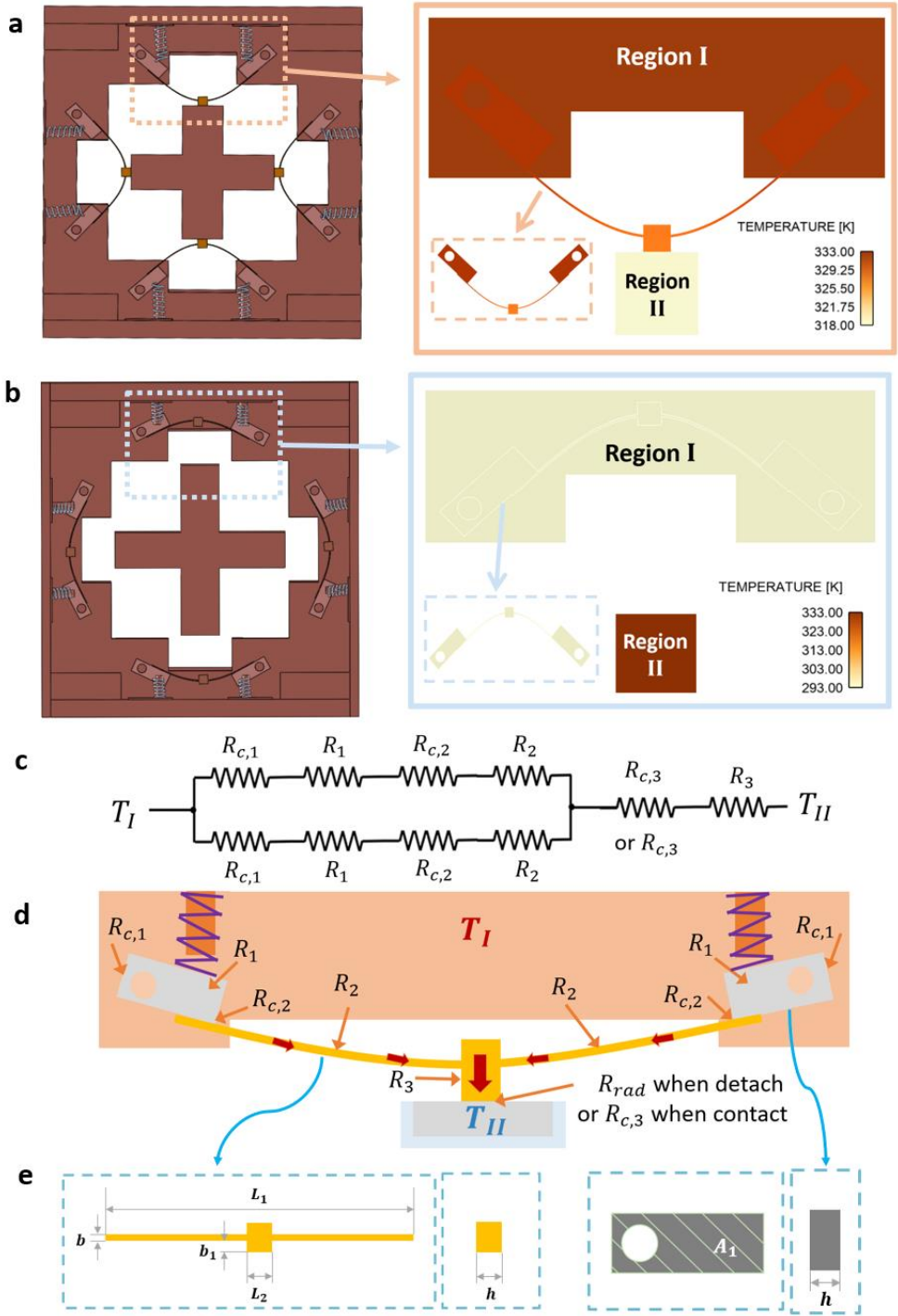

Figure 5. Thermal modeling of the thermal diode. FE simulation of temperature distributions in the (a) forward and (b) backward modes. (c) Equivalent 1D thermal circuit. (d) Resistive components in the forward and backward conduction paths. (e) Geometric description of the copper strip and central copper block, as well as geometry for rotational hinges. Here $R_i$ is the conduction resistance, $R_{c,i}$ is thermal contact resistance, and $R_{rad}$ is the radiative thermal resistance. $T_I$ and $T_{II}$ represent the temperatures in regions I and II, respectively.

To quantify nonreciprocal thermal transport, we define the thermal rectification ratio as:

$$\eta = \frac{q_f - q_b}{q_b} = \frac{R_b - R_f}{R_f} \tag{8}$$

where $q_f$ and $q_b$ are the steady-state heat fluxes in the forward and backward modes, respectively. This metric captures the degree of directionality in thermal conduction enabled by the diode structure.

After computing the thermal resistances from Equations (6) and (7), the rectification ratio $\eta$ can be calculated via Equation (8). Figure 6a presents how $\eta$ varies with two critical geometric parameters: the thickness of the copper strip and the width of the central copper block, while keeping $L_1$, $b_1$ and $h$ (Figure 5e), and copper properties constant. The primary findings of the effects of parameters on $\eta$ are as follows:

- Increasing the copper strip thickness $b$ reduces the conduction resistance $R_f$, improving heat transport in the forward mode (Figure 6c) and boosting $\eta$.
- Widening the central block marginally lowers $R_c$, but the impact on total resistance is less significant than that of the strip thickness.
- In the backward mode, $R_b$ decreases with increased radiative surface area $A_r$, as shown in Figure 6d.

The optimized design achieves a record-high thermal rectification ratio of ~936, as further validated experimentally in Figure 6b. The analysis informs the design tradeoffs between thermal response and physical dimensions, aiding optimization for high-performance rectification.

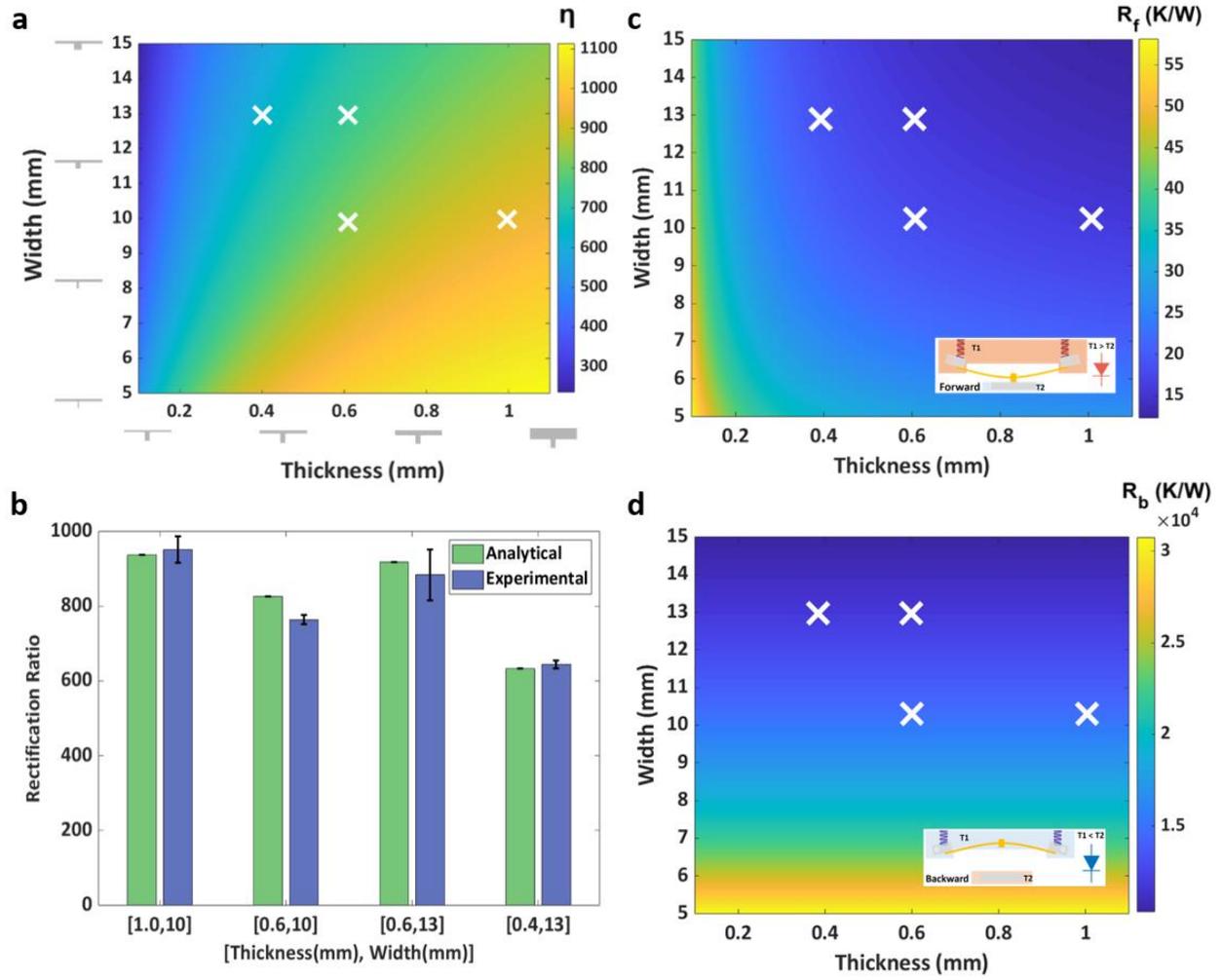

*Figure 6. Thermal performance of the diode. (a)Thermal rectification ratio $\eta$ as a function of copper strip thickness and central block width. Total thermal resistance versus central block width and strip thickness: (c) in forward mode ($R_f$) and (d) backward mode ($R_b$). The white crosses symbolize the experimental validation.*

## 6. Modular Design of Thermal Diodes

The thermal metamaterial illustrated in Figure 1a features a four-layer modular architecture, with each layer composed of four thermal diode units described in Section 3. Figure 7a shows the forward-mode heat transfer path in this modular configuration. As seen in the side-sectional view, the central block's height allows the direction of heat transfer near the heat sink to become perpendicular to the in-plane conduction direction of individual layers, enabling efficient vertical heat extraction.

In the backward mode (Figure 7b), when the central conductive block acts as a heat source, the air gaps between the central block and the diode units in each layer inhibit heat flow, effectively blocking conduction and reinforcing the rectifying behavior of the structure.

To model the stacked system, the thermal resistance of a single unit cell, defined previously in Equations (6) and (7), can be treated as an equivalent resistance $R_u$, as shown in the thermal circuit in Figure 7d. For one layer composed of four parallel units, the effective layer resistance is: $R_{layer} = R_u/4$ with $N$ such layers stacked vertically (in series), the total system resistance becomes:

$$R_{total} = \frac{1}{N} \cdot R_{layer} \tag{9}$$

Accordingly, for a four-layered modular system shown in Figure 7c, the overall thermal resistance in both forward and backward modes becomes one-sixteenth of that of a single unit, i.e., $R_{total} = R_u/16$. Importantly, since the thermal rectification ratio $\eta$ depends on the ratio of forward to backward resistance, it remains independent of the number of units:

$$\eta_{modular} = \eta_{unit} \tag{10}$$

Nonetheless, based on the general heat flow expression $q = \Delta T/R$, increasing the number of units leads to a larger effective cross-sectional area $A$, which reduces the thermal resistance and enhances absolute heat transfer, particularly in the forward mode. This improvement makes modular stacking an effective strategy for scaling the thermal diode's capacity while preserving its nonreciprocal behavior.

For experimental demonstration, we fabricated and tested only a single-layer architecture to clearly illustrate the thermomechanical deformation and snapping behavior of the diode unit (see Video S1). The four-layered modular structure is presented in Video S2 as an exploded-view rendering, serving to demonstrate the scalable assembly concept. The thermal performance of the modular system is evaluated through modeling as described above. In our experiments, the time required for the central copper block to fully contact the heat sink under heating was approximately 180 seconds, as shown in Supplementary Video S1.

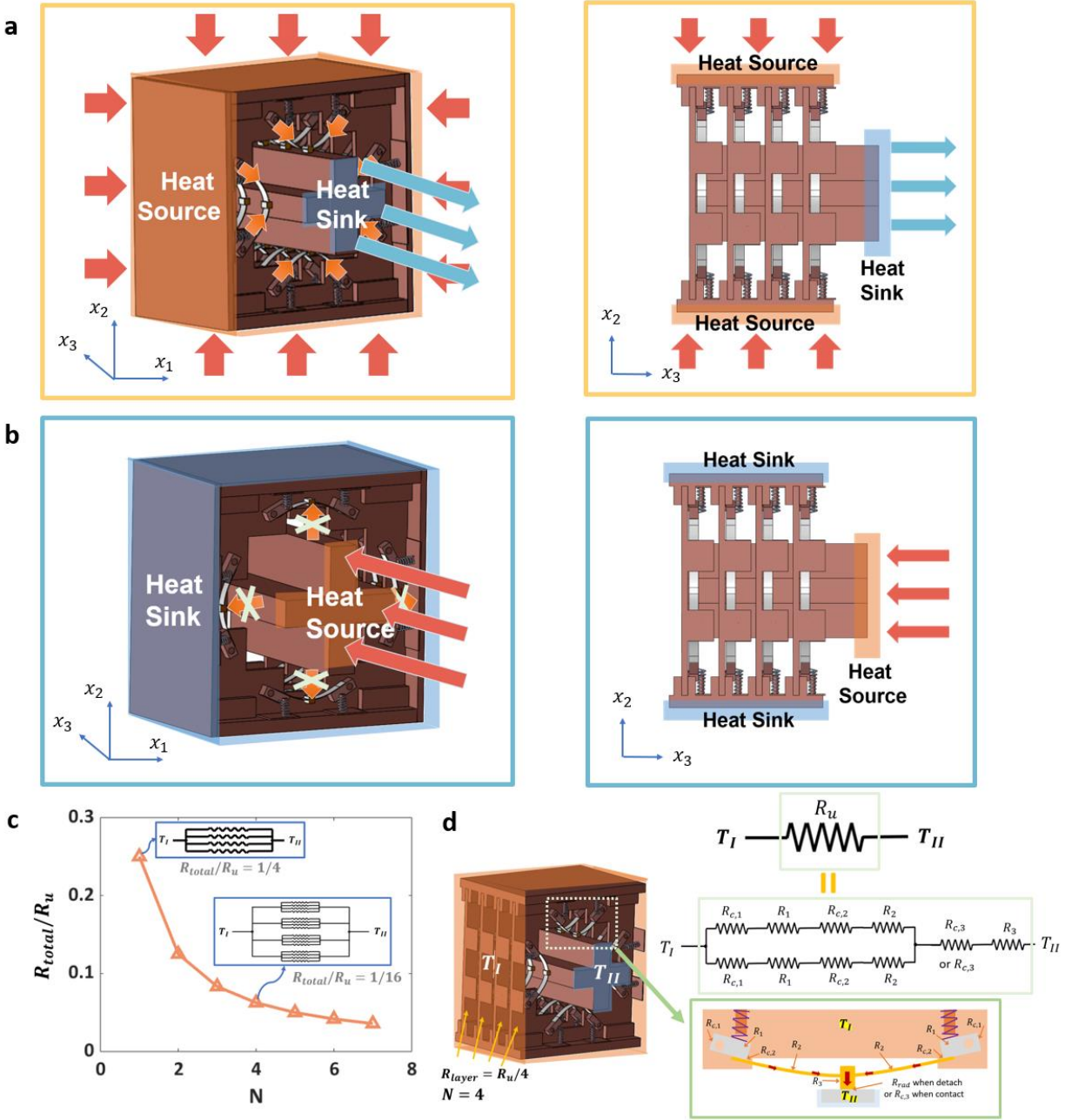

*Figure 7. Modular thermal diode design with enhanced directional heat transfer. (a) Forward operating mode: Oblique view showing modular stacking of four layers, each with four thermal diode units, and side-sectional view illustrating vertical heat flow from the central block to the heat sink. (b) Backward operating mode: Oblique view showing thermal isolation due to air gaps, and sectional view highlighting disruption of thermal pathways. (c) Schematic showing total thermal resistance reduction as a function of the number of stacked layers $N$. (d) Equivalent thermal circuit diagram of a single diode unit. Here, $R_u$ is the equivalent resistance of a single unit cell, $R_{layer}$ is the equivalent resistance of one layer, and $R_{total}$ is the total system resistance.*

## 7. Discussion

The thermomechanical metamaterial presented in this study represents a substantial advancement in the design of thermal diodes by integrating SMA actuation with structural bistability to achieve direction-dependent heat transport. The device operates by enabling mechanical contact in the forward mode to facilitate conduction, while maintaining separation in the reverse mode to suppress thermal transfer via radiation. As reported in Table 1, the diode achieves a rectification ratio of 936—the highest experimentally validated value among macroscale thermal diodes—demonstrating its effectiveness in producing strong thermal asymmetry under passive operation.

Unlike conventional thermal diodes, which typically exploit temperature-dependent material properties such as phase transitions[12,13], interfacial asymmetry[14], or phonon scattering[15,16], the present design employs a fundamentally distinct mechanism driven by temperature-induced mechanical reconfiguration. Prior approaches have faced critical limitations: phase-change and silicon-based diodes generally yield rectification ratios below 11[15,16] and are hindered by hysteresis and slow switching kinetics; droplet-based systems degrade over time due to flooding effects[13]; and graphene-based configurations lack geometric tunability and remain largely theoretical due to fabrication constraints[14]. The approach introduced here avoids these issues by realizing nonreciprocal heat flow through a mechanically stable architecture actuated by thermal input alone, without the need for continuous power or external control systems.

In addition to its high performance, the proposed thermal diode offers clear potential for application in advanced thermal management. Its ability to passively switch between thermally conductive and insulative states based on ambient temperature allows it to function as a thermal gate that responds directly to environmental conditions. This feature is particularly advantageous for systems that require spatially and temporally adaptive heat control, such as aerospace structures, battery enclosures, and thermal shielding in energy systems. As illustrated in Figure 8a–b, the diode's integration into a water pipe structure demonstrates its potential capacity to regulate heat exchange under varying ambient temperatures—promoting heat gain when external conditions are hot, and preventing heat loss from water when the surroundings are cold—thereby highlighting its utility in infrastructure and thermal packaging applications.

To assess thermal efficiency under ambient conditions, experiments were conducted with an insulating bubble wrap layer to suppress convective heat loss (Figure 8c). A detailed calculation of heat loss due to air convection is provided in Section S5 in the Supplementary Information. The results, summarized in Figure 8d, indicate that even under elevated convective coefficients, the heat loss remains below 20% of the input energy, ensuring that the thermal asymmetry is maintained without significant energy dissipation.

*Table 1. Comparative analysis of thermal rectification performance across multiple scales and mechanisms.*

| Rectification Ratio (η)* | Operating Mechanism | Scale | Reference |
|---|---|---|---|
| **Experimental-theoretical Validation** | | | |
| 936 | SMA-actuated bistable metamaterial | Macroscale | This work |
| 325 | Electric-field-enhanced phase-change | Macroscale | [33] |
| 2.02 | Tesla valve vapor channel | Macroscale | [34] |
| 1.77 | Fractal geometry with liquid metal interface | Macroscale | [35] |
| 0.14 | Nanostructured Si thermal nonlinearity | Microscale | [16] |
| **Simulation-theoretical Validation** | | | |
| 10000 | Near-field Si nanoparticle radiation | Nanoscale | [36] |
| 99 | Coupled nonlinear 1D lattices | Nanoscale | [37] |
| **Hybrid (Experimental and Simulation) Validation** | | | |
| 49 | Transient thermoelectric shielding/harvesting | Macroscale | [25] |
| 0.6 | Asymmetric TNM (copper/EPS) | Macroscale | [38] |
| 0.109 | Nano Thermal Mechanical gap modulation | Microscale | [39] |

*\* Revised thermal rectification ratios by Equation (8).*

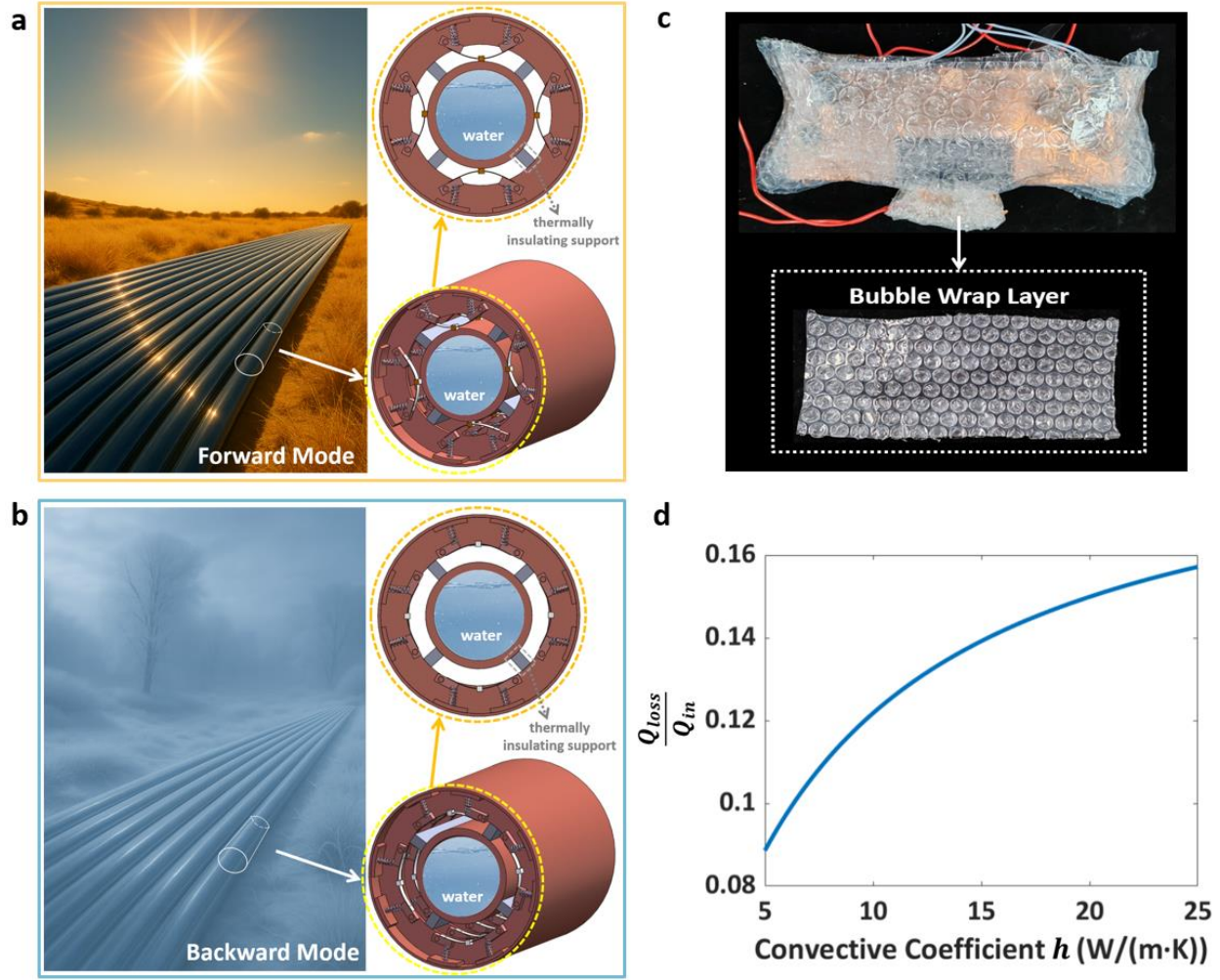

*Figure 8. Potential applications in water pipes under different temperatures outside: (a) High temperature outside. (b) Low temperature outside. (c) Thermal experimental setup with bubble wrap layers. (d) Ratio of heat loss $Q_{loss}$ to input energy $Q_{in}$ as a function of convective heat transfer coefficient $h$.*

## 8. Conclusions

In this work, we introduced a thermomechanical metamaterial-based thermal diode that achieves record-high rectification by coupling snap-through bistability of a column buckled copper beam with temperature-responsive actuation via SMAs. The device exploits a mechanically regulated switching mechanism to establish conductive contact in the forward mode and thermally insulating separation in the backward mode. This configuration enables directional, nonreciprocal heat transport with a measured rectification ratio of 936—the highest experimentally validated value reported at the macroscale.

The design further supports modular scalability through stacked architectures composed of discrete diode units. This modularity allows for increased thermal throughput while maintaining directional selectivity, making the system adaptable to a range of operating conditions and heat load requirements. Experimental demonstrations, such as the integration with temperature-adaptive water pipes, highlight the potential of the device for intelligent thermal regulation in built environments and thermal protection systems.

The thermal diode introduced in this work sets a new benchmark for nonreciprocal heat transfer in macroscale systems. Its integration of bistable mechanics and SMA-driven actuation enables directional heat transport with strong rectification and passive environmental responsiveness. By departing from traditional material-based mechanisms and embracing a thermomechanical paradigm, this design opens promising avenues for the development of intelligent, scalable, and application-ready thermal regulation technologies. Future work will focus on tuning actuation thresholds through material and structural design, improving thermal response times, and validating device stability under extended cyclic loading to further support its adoption in real-world systems.

## Acknowledgements

This research was supported by the Ministry of Science and Technology in China (Grant no. 2022YFE0129000 (J.J.)), the National Natural Science Foundation of China (Grant no. 12272225 (J.J.)), and the Research Incentive Program of Recruited Non-Chinese Foreign Faculty by Shanghai Jiao Tong University (J.J.). The authors also thank the members of S-Lab at the UM-SJTU Joint Institute for their valuable technical discussions.

## Conflict of interest

The authors declare no conflict of interest.

## Author Contributions

J.J. was responsible for the design and supervision of the research. Q.D. conceptualized the mechanism, designed the structures, developed analytical models, and carried out FE simulations. Y.W. and Q.Y. fabricated the devices and conducted experiments. Y.C. contributed to the three-dimensional model and supplementary videos. G.X., W.C., and Z.W. contributed to the establishment of the thermal mechanism. All authors collaboratively wrote the manuscript. A.N. contributed to the financial support for the project.

## Data Availability Statement

The data that support the findings of this study are available in the supplementary material of this article.